\documentclass[graybox,natbib,nosecnum]{svmult}
\bibpunct{(}{)}{;}{a}{}{,} % suppress commas between author-names and year

\pdfoutput=1   %forces use of pdflatex. Disable if you prefer to use .eps or .ps figures.
\usepackage{mathptmx}       % selects Times Roman as basic font
\usepackage{helvet}         % selects Helvetica as sans-serif font
\usepackage{courier}        % selects Courier as typewriter font
\usepackage{type1cm}        % activate if the above 3 fonts are
\usepackage{makeidx}         % allows index generation
\usepackage{graphicx}        % standard LaTeX graphics tool
\usepackage{multicol}        % used for the two-column index
\usepackage[bottom]{footmisc}% places footnotes at page bottom
\usepackage[normalem]{ulem}     % for strike-through of text with \sout{}
\usepackage{hyperref}  %for hyperlinks

\usepackage{soul}   % for high-lighting of text
\makeindex             % used for the subject index
\begin{document}

\title*{Extrasolar Kuiper Belts}
\author{Mark C. Wyatt}
\institute{Mark C. Wyatt \at Institute of Astronomy, University of Cambridge, Madingley Road, Cambridge CB3 0HA, United Kingdom, \email{wyatt@ast.cam.ac.uk}}
\maketitle

\abstract{Extrasolar debris disks are the dust disks found around nearby main sequence stars arising from the break-up of asteroids and 
comets orbiting the stars.
Far-IR surveys (e.g., with {\it Herschel}) showed that $\sim20$\% of stars host detectable dust levels.
While dust temperatures suggest a location at 10s of au comparable with our Kuiper belt,
orders of magnitude more dust is required implying a planetesimal population
more comparable with the primordial Kuiper belt.
High resolution imaging (e.g., with ALMA) has mapped the nearest and brightest disks, providing evidence for structures
shaped by an underlying planetary system.
Some of these are analogous to structures in our own Kuiper belt (e.g., the hot and cold classical, resonant, scattered disk and
cometary populations), while others have no Solar System counterpart.
CO gas is seen in some debris disks, and inferred to originate in the destruction of
planetesimals with a similar volatile-rich composition to Solar System comets.
This chapter reviews our understanding of extrasolar Kuiper belts and of how our own Kuiper belt compares with
those of neighbouring stars.}

%\begin{keywords}[Keywords:]
%  Extrasolar planetary systems, debris disks, exocomets, planetary system evolution, planetary system dynamics
%\end{keywords}

%%%%%%%%%%%%%%%%%%%%%%%%%%%%%%%%%%%%%%%%%%%%%%%%%%%%%
%%%%%%%%%%%%%%%%%%%%%%%%%%%%%%%%%%%%%%%%%%%%%%%%%%%%%
\section{Introduction}
\label{s:introduction}
As the other chapters in this book testify, studies of the Kuiper belt continue to play a
key role in shaping our understanding of the structure and history of the Solar System.
This chapter aims to provide some context to this relatively detailed information on our own
planetary system, by considering how this compares to the planetary systems of other stars,
in particular with regard to the component of their debris disks
which might be considered analogous to the Kuiper belt.
Such comparisons help to ascertain whether our system is typical or atypical, either
in terms of its current architecture, or in terms of its past history. 
They also allow a deeper understanding of processes inferred to have occurred in our own history,
since those processes may be ongoing for some stars, furthermore providing evidence for how these
play out in different environments.

By now it is clear that our Solar System is just one of many planetary systems, with observations
of nearby stars showing that approximately half of stars have a planetary system in which at least
one of its planets may be detected in current surveys \citep{Winn2015}.
However, this seemingly abundant information is restricted to the planets that reside close to their
stars (within a few au), and information about the outer regions of planetary systems remains
scarce.
Indeed just a few percent of stars have planets $\gg 5$\,au detectable by direct imaging (which means at
least a few times more massive than Jupiter), although microlensing surveys hint that Neptune-analogues
may be more common.

In contrast, our understanding of the dust content of outer planetary systems is relatively well advanced
\citep[e.g.,][]{Wyatt2008, Hughes2018}.
It is over 30 years since the far-IR satellite {\it IRAS} first discovered circumstellar dust orbiting nearby main sequence
stars \citep{Aumann1984}.
Because this dust is short-lived it was recognised that it must be continually replenished from
larger planetesimals situated in a source region that was inferred from the dust temperature to be
at 10s of au \citep{Backman1993}, a conclusion that was reinforced by imaging of the dust structures \citep{Smith1984}.
Stars with such dust have been called Vega-type (after the first discovery), and the circumstellar dust
is referred to as a debris disk
\footnote{Here {\it debris} refers to the debris left over after planet formation, which
applies to any circumstellar material that is not a planet.
This could include a dust or gas component left over from the protoplanetary disk,
but also planetesimals and the dust and gas resulting from their destruction.},
and often interpreted to arise from an extrasolar analogue to the
Solar System's Kuiper belt \citep[e.g.,][]{Wyatt2003a, MoroMartin2008}. 

The extent to which that analogy is appropriate
remains open for interpretation, but what is clear is that $\sim 20$\% of
stars have dust, and so presumably also planetesimals, at somewhere around 20-150\,au from their
stars \citep{Wyatt2008, Hughes2018}.
Given the aforementioned difficulty of detecting planets in these outer regions, it is usually uncertain
whether this dust lies at the outer edge of a planetary system or whether there are other similarities
with our own Kuiper belt, but there are clues in the dust structures.
These often reinforce the Solar System analogy, with the caveat that there is some anthropocentric
bias in this interpretation.

This chapter starts in \S \ref{s:observations} by summarising the various observational methods
used to obtain information about these putative extrasolar Kuiper belts, then in
\S \ref{s:perspective} considers how our Solar System fits within this context. 
More detail is given in \S \ref{s:properties} about specific aspects about the physical and
dynamical properties of the extrasolar systems and their comparison with the Solar System,
before concluding in \S \ref{s:conclusions}.

%%%%%%%%%%%%%%%%%%%%%%%%%%%%%%%%%%%%%%%%%%%%%%%%%%%%%
%%%%%%%%%%%%%%%%%%%%%%%%%%%%%%%%%%%%%%%%%%%%%%%%%%%%%
\section{Extrasolar Kuiper Belt Observations}
\label{s:observations}
As noted in \S \ref{s:introduction}, most of the information about extrasolar Kuiper belts comes from observations
of dust created in the destruction of their planetesimals.
These observations can be roughly split into those used to discover the presence of dust using photometry (discussed
in \S \ref{ss:discovery}) and those that provide more detailed characterisation through high resolution
imaging (discussed in \S \ref{ss:characterisation}).
There is also an emerging area of observations of gas in the systems (see \S \ref{ss:gas}).

%%%%%%%%%%%%%%%%%%%%%%%%%%%%%%%%%%%%%%%%%%%%%%%%%%%%%
\subsection{Discovery: Photometry and SED Fitting}
\label{ss:discovery}
Debris disk dust is most readily detected from its thermal emission which manifests itself as the star appearing brighter
than expected from purely photospheric emission at wavelengths that depend on the dust temperature,
although care must be taken to ensure that potential extragalactic contamination has been removed from
the photometry \citep[e.g.,][]{Kennedy2012, Gaspar2014}.
The emission from dust created at 10s of au peaks at far-IR wavelengths and that is where the majority of debris
disks have been discovered.
The most recent surveys were those undertaken by {\it Herschel} \citep{Eiroa2013, Matthews2014, Hughes2018},
including an unbiased survey of the nearest several hundred stars in which dust emission was detected around
20\% of stars \citep{Thureau2014, Sibthorpe2018}.

\begin{figure}
\includegraphics[scale=0.51]{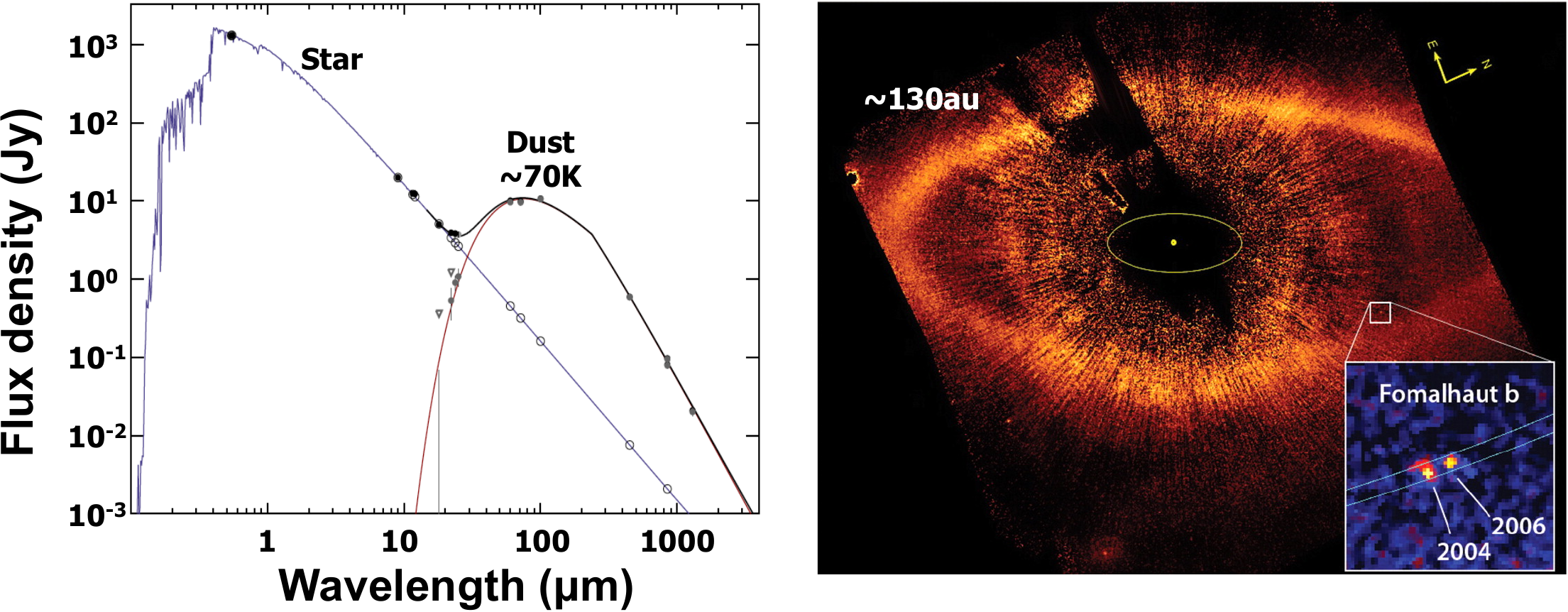}
\caption{Observations of the nearby (7.8\,pc) 400\,Myr A-type main sequence star Fomalhaut.
{\it (Left)} Photometric observations show that the spectral energy distribution is comprised of
two temperature components, one from the star and another from cold dust at $\sim 70$\,K.
{\it (Right)} Optical images in which the stellar emission has been subtracted using coronagraphic techniques
to reveal scattered light from circumstellar dust that is distributed in a narrow ring at $\sim 130$\,au
from the star \citep{Kalas2008}.
The centre of this ring is offset from the star implying an eccentricity of $\sim 0.11$, and a planet-like object 
(Fomalhaut-b) is seen close to the inner edge of the ring.}
\label{fig:1}
\end{figure}

These photometric surveys are supplemented by photometry at mid-IR and sub-mm wavelengths to build up
the spectral energy distribution (SED) of the dust emission, which generally shows that the emission is dominated
by a single temperature and so dust at a single distance (see e.g. Fig.~\ref{fig:1}). 
Given the luminosity of the star this can be translated into a black body distance (i.e., the distance at which
the dust would be if it behaved like a black body), although it is recognised that the small dust that dominates
is an inefficient emitter and so this is likely an underestimate of the true location of the dust by a factor that
can be several depending on the star \citep{Booth2013, Pawellek2014}.
In addition to its black body radius $r_{\rm bb}$, the disk's SED is also characterised by its fractional
luminosity $f$ (i.e., the luminosity of the dust
divided by that of the star), with values of $10^{-6}$ to $10^{-3}$ typical for known disks.

For some systems the SED has been characterised in greater detail, in which case more realistic modelling
is performed to constrain the dust size distribution and composition \citep[e.g.,][]{Olofsson2012, Lebreton2012}.
However, the simple shape of the spectrum means that there is still value in considering the disks in terms
of their two observable parameters ($r_{\rm bb}$ and $f$), particularly because there are a number of
observational biases that can be readily understood within this context \citep{Wyatt2008}.
Nevertheless, some systems have an SED that is best characterised as having two temperatures \citep{Chen2006, Kennedy2014}.
In the context of an extrasolar Kuiper belt interpretation, this means that there is an additional
warm component of emission that might be analogous to the Solar System's zodiacal cloud.
However, there remains debate about whether this component originates in an extrasolar asteroid belt
analogue \citep{Su2013}, exocomets scattered in from the extrasolar Kuiper belt \citep{Wyatt2017}, or is a transient 
dust component perhaps caused by a recent collision \citep{Jackson2012}.

%%%%%%%%%%%%%%%%%%%%%%%%%%%%%%%%%%%%%%%%%%%%%%%%%%%%%
\subsection{Characterisation: Imaging}
\label{ss:characterisation}
Debris disks around the nearest stars have a spatial extent of $>1$\,arcsec which means that these can be readily
resolved.
The low resolution of far-IR instrumentation precludes this except in a few cases \citep[e.g.,][]{Acke2012}, 
but this technique has been
successful {at optical and near-IR wavelengths} with {\it HST}
(see Fig.~\ref{fig:1}), and more recently SPHERE on VLT and GPI on Gemini,
where starlight scattered by the smallest sub-$\mu$m
dust can be imaged \citep[e.g.,][]{Schneider2014}.
In the sub-mm with ALMA (Atacama Large Millimeter/submillimeter Array),
thermal emission from mm to cm-sized grains can be
imaged as well (see Fig.~\ref{fig:5}).
Given the size of dust dominating the emission at the different wavelengths, it is generally assumed that
sub-mm observations trace the distribution of the parent planetesimals, while the picture is more complicated
at optical wavelengths because the orbits of small dust grains are significantly modified by radiation forces.
Indeed, different radial and even nonaxisymmetric structures are seen in multiwavelength imaging of the
same disk providing a valuable diagnostic tool for the underlying dynamics.

Such imaging provided the first evidence that the dust is configured in a disk \citep[rather than a spherical distribution,][]{Smith1984},
and has allowed direct measurement of the radial location of the dust, its radial and vertical extents and the presence of gaps
(see \S \ref{ss:radial}),
as well as providing evidence for asymmetries in the form of clumps, eccentricities and warps (see \S \ref{ss:structures}).
All of these give vital clues to the underlying dynamics and the existence of planets.

%%%%%%%%%%%%%%%%%%%%%%%%%%%%%%%%%%%%%%%%%%%%%%%%%%%%%
\subsection{Gas}
\label{ss:gas}
Debris disks used to be considered as gas-free, in contrast to the protoplanetary disks found around
young stars ($<10$\,Myr). 
Recent discoveries with ALMA show that this can no longer be considered the case, since there are now of order
10 main sequence stars known to exhibit CO gas emission.
Many of the first gas detections are around the more massive A-type stars \citep{Kospal2013, Greaves2016, Moor2017},
though there are a number of observational biases that facilitate the detection of gas around early-type stars,
and gas is now seen around later spectral types as well \citep{Marino2016, Matra2019a}.
Most of these observations can be explained within the context of a model in which the gas in debris disks,
like the dust, is the product of of the destruction of planetesimals that are then required to be volatile-rich
\citep{Kral2017}.
An open question is whether the youngest ($<50$\,Myr) stars retain gas from the protoplanetary disk phase
\citep{Kospal2013, Kral2018}.
Observations of debris disk gas not only allow the mass and composition of gas to be measured (see \S \ref{ss:composition}),
but also spatially resolved
with additional kinematic information, resulting in a potentially powerful probe of the disk structure and dynamics
\citep[see \S \ref{ss:structures},][]{Dent2014}.

%%%%%%%%%%%%%%%%%%%%%%%%%%%%%%%%%%%%%%%%%%%%%%%%%%%%%
%%%%%%%%%%%%%%%%%%%%%%%%%%%%%%%%%%%%%%%%%%%%%%%%%%%%%
\section{An Extrasolar Perspective of the Kuiper Belt}
\label{s:perspective}
\begin{figure}
\includegraphics[scale=0.49]{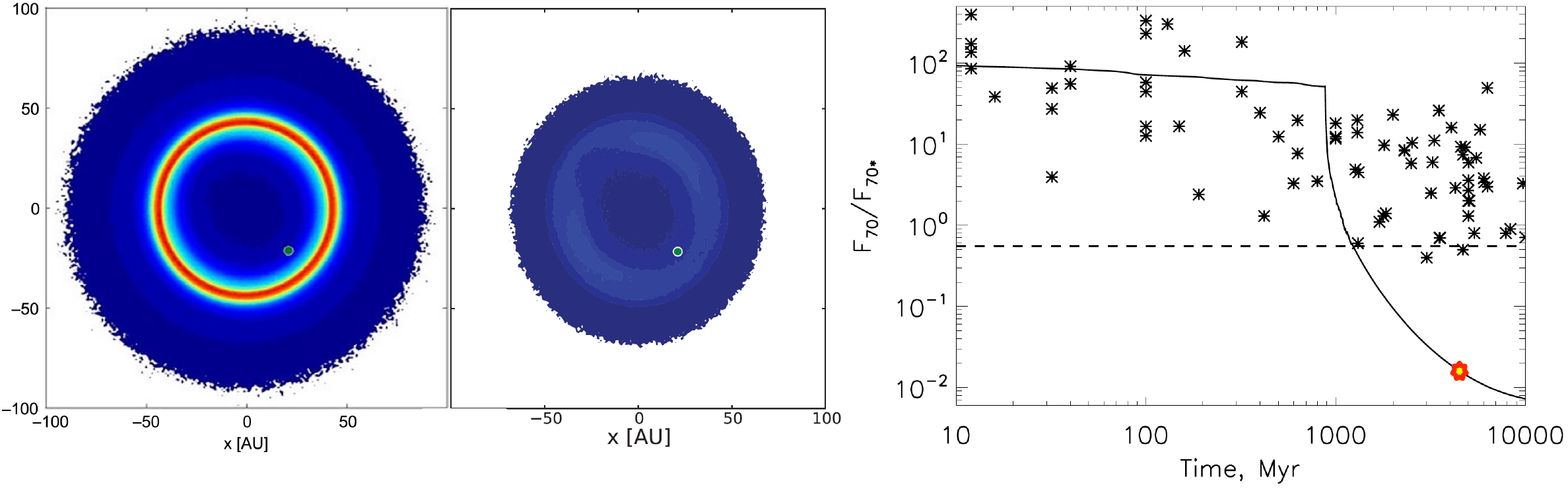}
\caption{The Solar System's Kuiper belt.
{\it (Left)} Face-on images of the number density distribution of Kuiper belt objects that have been debiased from the
observed distributions.
To the left is shown the full Kuiper belt which appears as a narrow ring dominated by the classical Kuiper belt
objects \citep{Matthews2016}, and to the right the contribution of resonant Kuiper belt objects that shows two clumps
at $\pm 90^\circ$ from Neptune (shown with a circle) due to the Plutino population \citep{Lawler2014}.
{\it (Right)} The evolution of the 70\,$\mu$m emission from dust in the Kuiper belt (shown as its level compared with
that from the Sun) expected in the model of \citet{Gomes2005}, and compared with the dust emission from nearby
Sun-like stars \citep{Booth2009}.
In this model the massive primordial Kuiper belt is comparable in brightness to those observed around nearby stars
(shown with asterisks),
but is depleted to a non-detectable level at $\sim 800$\,Myr by the Late Heavy Bombardment.
}
\label{fig:2}
\end{figure}

It is important to note that the Solar System's debris disk is fainter than any known debris disk around another star.
If one of the stars in the debris disk surveys hosted an exact replica of the Solar System, it would only have
been possible to detect the photospheric emission from the star and not that of circumstellar dust (see Figs.~\ref{fig:2}
and \ref{fig:3}).
The challenge with detecting a true Kuiper belt analogue lies in the fact that its thermal emission is outshined by more
than an order of magnitude at all wavelengths by the Sun.

In fact the level of thermal emission from dust in the Kuiper belt is not well known.
Its emission has not been detected since it is masked by that of the zodiacal cloud which is much closer to the
Earth, although the far-IR all-sky surveys like {\it IRAS}, {\it COBE} and {\it Planck}
do provide upper limits on the level of emission
that can be present \citep[e.g.,][]{Backman1995}.
Direct detection of dust grains in the outer Solar System gives an indication of the number of dust grains present
\citep{Piquette2019}, but this must
be combined with dynamical models for dust production and evolution to understand the whole dust structure \citep{Vitense2014}.
In doing so, the best estimate is that Kuiper belt emission peaks at $\sim 70$\,$\mu$m at a flux 1\% that of the Sun \citep{Vitense2012}. 
Even for bright stars their far-IR fluxes cannot be predicted with such accuracy, which combined with calibration uncertainties
in photometric measurements, means that $\gg 10$\% excesses are required for a confident detection.

If it were possible to resolve the emission from Kuiper belt dust, the structures that would be observed
at long wavelengths would resemble the structures known to be present in the distribution of Kuiper belt objects
\citep[e.g.,][]{Lawler2014};
i.e., there would be a prominent ring at $\sim 40$\,au from the classical Kuiper belt
(see Fig.~\ref{fig:2} left),
with some non-axisymmetric structure caused by the resonant Kuiper Belt Objects (see Fig.~\ref{fig:2} middle),
and with emission extending out to larger radii from the scattered disk (which is present but hardly noticeable in
Fig.~\ref{fig:2} left).
Given the low density of the dust distribution, the smallest dust (as traced by the shorter far-IR wavelengths) would
migrate in by Poynting-Robertson drag and so fill in the 40\,au hole. 
However, very little dust would make it past Jupiter or even Saturn, meaning that this tenuous dust distribution
would have drops in density associated with these planets \citep{Liou1999, MoroMartin2002}.

There are, however, indications that the Kuiper belt was more massive in the past by several orders of
magnitude.
At such an epoch there would have been correspondingly more dust, resulting in detectable levels of
emission \citep[see Fig.~\ref{fig:2} right,][]{Booth2009}.
Thus the known debris disks could be analogues to the primordial Kuiper belt, and thus representative of
systems that either never will, or have yet to undergo, large-scale instabilities.
Note though that the exact timing of the depletion in the Solar System remains uncertain \citep[e.g.,][]{Gomes2005, Morbidelli2018}.

%%%%%%%%%%%%%%%%%%%%%%%%%%%%%%%%%%%%%%%%%%%%%%%%%%%%%
%%%%%%%%%%%%%%%%%%%%%%%%%%%%%%%%%%%%%%%%%%%%%%%%%%%%%
\section{Extrasolar Kuiper Belt Properties (and Comparison with Solar System)}
\label{s:properties}

%%%%%%%%%%%%%%%%%%%%%%%%%%%%%%%%%%%%%%%%%%%%%%%%%%%%%
\subsection{Mass and Radius Distribution}
\label{ss:mass}
Far-IR surveys from the past few decades have provided large quantities of information
on the incidence of debris disks, as measured by the excess emission above photospheric
levels at different wavelengths (typically at 24, 70, 100 and 160\,$\mu$m).
Surveys have been carried out for nearby stars with a variety of spectral types, distances and
ages.
While main sequence star ages are usually quite inaccurate, there is sufficient information, particularly
when comparing associations of young stars for which ages can be better determined, to show that
disks tend to get fainter with age \citep{Rieke2005, Su2006}.
As such the surveys are usually interpreted within the context of population models that take
collisional erosion into account (see \S \ref{ss:evolution}).

\begin{figure}
\hspace{0.5in}
\includegraphics[scale=0.75]{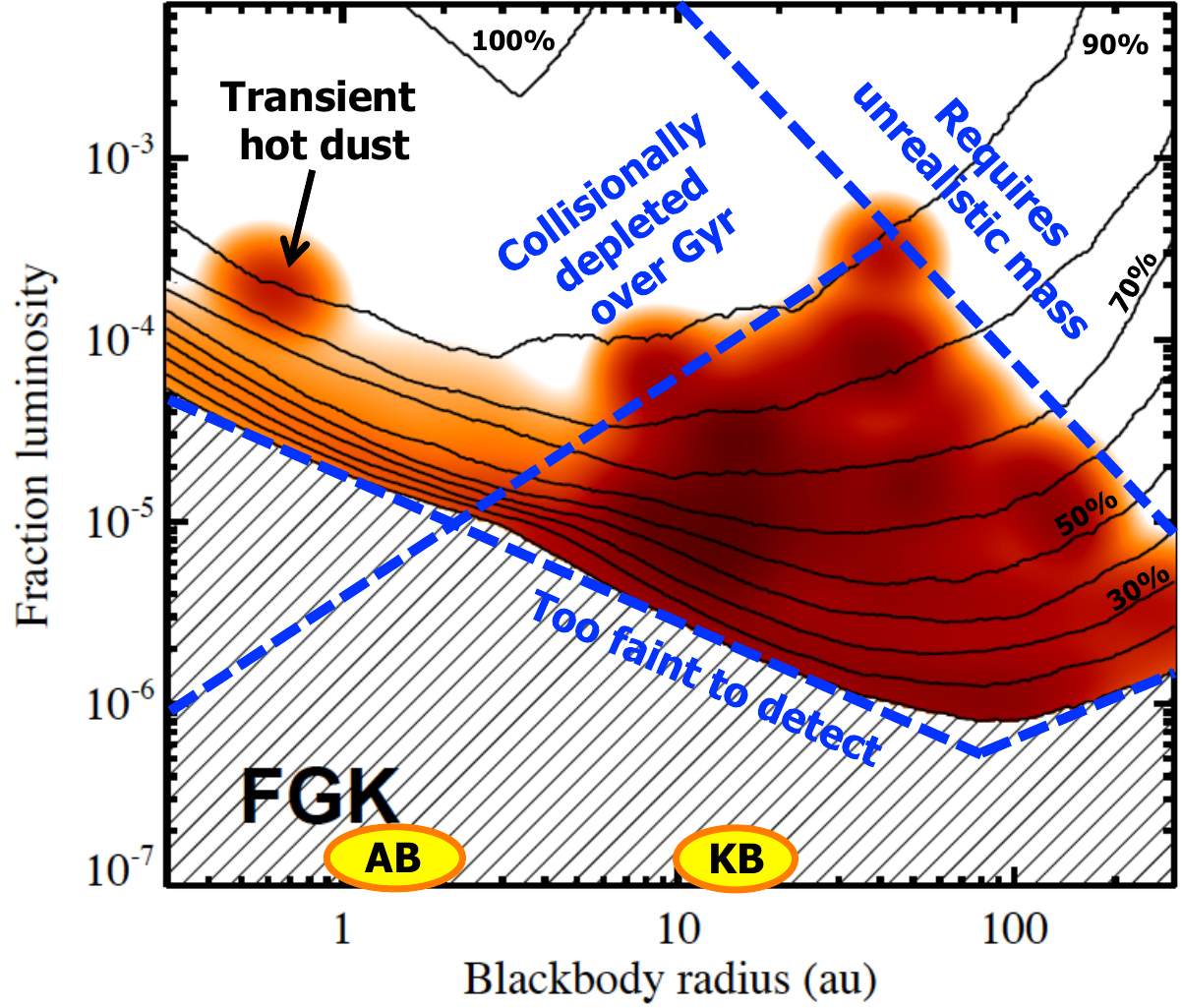}
\caption{Distribution of debris disk luminosities and radii \citep[figure adapted from][]{Sibthorpe2018}.
The colour scale shows the fraction of a sample of $\sim 300$ Sun-like (FGK) stars from \citet{Phillips2010}
that have dust emission as a function of their measured fractional luminosities and black body radii.
The approximate locations of asteroid belt and Kuiper belt dust are shown in yellow, noting that the black body
radius is a factor of a few smaller than the true radius because dust emits hotter than black body.
The contours show the fraction of stars in the sample for which dust emission could have been detected,
which is used to correct the observed incidence of debris to get the fraction plotted in the colour scale.
The blue dashed lines and annotation explain why the known disks (which are present around $\sim 20$\% of stars)
lie mostly in a well defined region of parameter space.
}
\label{fig:3}
\end{figure}

Fig.~\ref{fig:3} summarises the state of our knowledge of the debris disk population seen around nearby
Sun-like (i.e., F, G or K-type) stars. 
For each of the $\sim 300$ stars in the sample, the various observations (be they detections or non-detections
of dust emission) have been combined to determine the best fit fractional luminosity and black body radius to their disk
(see Fig.~\ref{fig:1}),
or the upper limit on the dust luminosity for a given temperature to allow a non-detection.
Particularly where disks are only detected at a single wavelength the fitted parameters can have significant
uncertainty, and so the disks are placed on this plot accordingly
\footnote{More specifically, uncertainties in the photometric data mean that analysis of the
spectral energy distribution can only determine a probability that the disk is present in a given
part of fractional luminosity vs black body radius parameter space.}, and the colour scale shows the fraction of stars
in the sample predicted to have a disk in different regions of parameter space \citep[for more details on
the method, see][]{Sibthorpe2018}.
Since observational biases mean that disks of a given level could only have been detected around a fraction
of the stars in the sample as shown in the contours \citep{Wyatt2008}, this
is taken into account to get an estimate of the underlying distribution.

As noted in the figure, the distribution of known disks is bounded on the bottom edge by an
observational limitation, while the top edges are set by physical considerations.
For disks close to the star ($<40$\,au in black body radius) the maximum fractional luminosity is set by collisional erosion,
since most nearby stars have the same roughly Gyr age and so their disks would be expected to have undergone similar
levels of collisional processing and would have undergone more depletion the closer they are to the star
\citep{Wyatt2007b}.
Some disks may exhibit higher fractional luminosities if there happens to be a young star in the sample, or if there is
a transient dust component \citep{Wyatt2007a}. 
For disks further from the star, the upper edge is set by a constraint on the mass required to 
produce the given fractional luminosity. 

While Fig.~\ref{fig:3} gives a quick visual impression of the distribution of debris disk properties, population models
that also address the age dependence of the phenomenon can be used to get a more accurate parameterisation
for the distributions of initial masses and radii of the disks.
Such models show that it is possible to fit the observations by assuming that all stars started out with a planetesimal
belt, with a radius and initial mass chosen from given distributions, that mass then evolving with time due to collisional erosion
\citep{Wyatt2007b, Lohne2008, Gaspar2013, Sibthorpe2018}, although it is possible that a few
of the brightest disks may require unrealistic initial masses \citep[$\sim 1000M_\oplus$,][]{Krivov2018}.
There is evidence that the disks of A-type stars tend to be brighter than their Sun-like star counterparts \citep{Greaves2003},
even accounting for the various biases (such as the on average younger ages of the more massive stars), which may be because of a difference
in their planetesimal belt properties (e.g., in having more massive disks, or one made of smaller planetesimals, or lower levels of
stirring).
Given that many of the well known bright debris disks discussed in this review are found around A stars (like $\beta$ Pic),
it may be worth bearing this caveat in mind when making comparisons with the Solar System.
There are few disks detected around the less massive M stars, but this is likely an observational bias and there is no
evidence as yet to say that their disks are any different to those of Sun-like stars \citep[e.g.,][]{Morey2014}.

Despite the success of these models, it is worth noting that they cannot make any strong conclusions about the 80\% of stars
that do not have detectable disks, since these would be based on assumptions about the shape of the distributions.
For example, a common assumption is that the mass distribution is log-normal, with the non-detections arising from 
close-in disks that undergo rapid depletion and so are not detectable in the nearby star population, but it could be that the
mass distribution is bimodal, with 80\% of stars having no disk.
Also, more recently a correlation has been found between planetesimal belt radii determined from
high resolution imaging of the sub-mm dust emission and the luminosity of the star \citep{Matra2018b}.
If this holds up as more disks are resolved, this would suggest a preferential location for the formation of
planetesimal belts that could be linked to ice-lines in the protoplanetary disk for example, which would provide an
additional constraint on the population models (as well as significant insight into the origin of the belts).

Fig.~\ref{fig:3} also includes the location of the Solar System's Kuiper belt and Asteroid belt
(see \S \ref{s:perspective}).
This shows that roughly 20\% of stars have disks that are comparable in radius to the Kuiper belt,
in having dust at temperatures that give a black body radius of 10-100\,au, but that have at least
an order of magnitude more dust at fractional luminosities of $10^{-6}$ to $10^{-3}$.
For the reasons noted above it is not possible to quantify where our own Kuiper belt fits into the
distribution.

%%%%%%%%%%%%%%%%%%%%%%%%%%%%%%%%%%%%%%%%%%%%%%%%%%%%%
\subsection{Evolution: Collisional vs Dynamical Erosion}
\label{ss:evolution}
As noted in \S \ref{ss:mass}, the evidence that there is any evolution in debris disk properties comes
not from seeing those changes in any one system, but rather from seeing the change in properties for samples
of stars with different ages.
Inevitably then there remains uncertainty in how individual disks evolve.
However, since it is necessary for collisions to occur to replenish the dust that is observed, and that
dust is lost from the system relatively quickly (compared with the age of the star), some level of collisional
erosion is unavoidable.
That evolution has been studied extensively both analytically and numerically, in ways that have also been applied
to small body populations in the Solar System.
In the Solar System evidence for past collisional evolution can be found in the detailed shape of the asteroid and
Kuiper belt size distributions \citep[e.g.,][]{Bottke2005, Fraser2009, Singer2019}, whereas in extrasolar systems the evolution of the
size distribution in debris disks
is manifested in the evolution of disk brightness which must be inferred from the population studies.

There are several unknowns in the problem, such as the initial size distribution of the planetesimals,
their level of stirring (i.e., their collision velocity), and their size-dependent dispersal threshold. 
With these factors known, the collisional evolution would be well understood \citep[e.g.,][]{Wyatt2011},
and when coupled with models for dust optical properties that determine both its emission properties and loss due to
radiation forces, this can be readily translated into an evolution in the disk's observables.
In principle this means that these fundamental physical parameters can be extracted from the observations,
although in practise there are degeneracies between the parameters which make it hard to extract unambiguous
information, other than that the vast majority of observations are consistent with purely steady state
collisional erosion.

Indeed it is easier to identify the disk properties or processes that are incompatible with the observations,
at least on a population level.
For example, a slow decline in disk incidence with age that is consistent with collisional erosion means
that large scale dynamical depletions at 100s of Myr must be relatively rare \citep{Booth2009}, otherwise there would be
fewer bright disks at several Gyr ages.
Thus either the Kuiper belt is anomalous in this respect \citep[i.e., if its depletion is responsible for the
Late Heavy Bombardment,][]{Gomes2005}, or the depletion actually occurred much earlier on \citep{Morbidelli2018}.
It is also possible to conclude that the majority of debris disks are not broad planetesimal disks \citep[even if each is only
bright at one radius at any given time, e.g.,][]{Kenyon2010},
since broad disks are hot early on and cold later on in a manner that is incompatible
with the observed 24 and 70\,$\mu$m brightness evolution \citep[][see also \S \ref{ss:radial}]{Kennedy2010}.

However, for comparison with the Kuiper belt it is interesting to consider what has been concluded about the 
largest planetesimals in debris disks and the origin and level of stirring within them.
While observations are only sensitive to dust up to $\sim $cm-size, the short collisional lifetime of such dust means that
it must be continually replenished.
Extrapolating the size distribution and equating the resulting collisional lifetime with age of the star shows that
debris disks are replenished by planetesimals at least several km in size \citep{Wyatt2002}, with population models
pointing to sizes up to 100\,km \citep{Gaspar2013}.
It has been suggested that some disks may be comprised of much smaller planetesimals, say 10-100\,m, that are colliding
at very low velocity so that the collisions are barely disruptive \citep{Heng2010, Krivov2013}, explaining the lack of small
grains inferred to some systems \citep{Eiroa2011}, although chance alignment of background galaxies could be confusing the
interpretation of these stars \citep{Gaspar2014}.
Detailed observations of the halo of small grains seen to extend beyond a debris disk due to radiation pressure
has the potential to constrain the level of stirring, with eccentricities inferred to be $<1$\%
in one system \citep{Thebault2008}.
For some disks the vertical structure has been measured with inclinations inferred to be of order a few percent
\citep[][]{Matra2019b, Daley2019}.

The origin of the stirring is not well constrained.
It could arise from large bodies embedded in the disk that excite relative velocities to around their escape
velocity \citep{Kenyon2010}.
However, the growth of Pluto-sized objects at 100s of au in the disk takes too long to explain large disks
seen around Gyr-old stars.
One possibility is that these form directly from the protoplanetary disk (rather than from the collisional growth
of smaller planetesimals), or it could be that a massive disk of $\sim 200$\,km planetesimals can stir itself
without the need for larger bodies \citep{Krivov2018b}.
If the belts do mark the outer edge of the planetary system then those planets could stir the disk through their
gravitational perturbations.
This could be through overlapping resonances from a planet at the inner edge \citep{Quillen2006}, but
could also arise from the secular perturbations of a more distant eccentric or inclined planet \citep{Mustill2009}.
It is also possible that the planetesimals were born on stirred orbits, e.g. if these formed in the early
phases of the protoplanetary disk \citep{Booth2016}.

To illustrate the range of interpretations for debris disks that do not encompass a traditional Kuiper belt-like
analogy, note that in one model the planetesimals form during protoplanetary disk dispersal from the
dust that is swept up by a photoevaporating gas disk \citep{Carrera2017}.
While such a model does not recover the observed distribution of debris disk radii, it is worth bearing in mind that
our understanding of protoplanetary disk dispersal processes is incomplete,
in particular regarding the fate of the $>1M_\oplus$ of mm-sized dust that
resides in the outer regions of the system \citep{Wyatt2015}.
As such it is not inconceivable that our understanding of the origin and evolution of debris disks could need
revision.

%%%%%%%%%%%%%%%%%%%%%%%%%%%%%%%%%%%%%%%%%%%%%%%%%%%%%
\subsection{Planetesimal Composition}
\label{ss:composition}
A standard way of determining the composition of circumstellar material is to look for solid state
features in the emission spectrum, such as the silicate features near 10\,$\mu$m.
This has been successful for the few systems for which warm dust is present \citep[e.g.,][]{Lisse2009},
however for the majority that are too cold to have significant mid-IR emission, this leaves far-IR features
of which only a few were measured with {\it Herschel}.
Nevertheless, when applied to the disk of $\beta$~Pic, the 69\,$\mu$m feature was used to show that 3.6\%
of the dust mass is in crystalline olivine grains, moreover constraining them to be relatively Mg-rich
\citep{deVries2012}.

As noted in \S \ref{ss:gas}, a new technique has been developed to probe the volatile content of the planetesimals 
by studying gas in the systems.
Molecular gas has a short and well defined lifetime in optically thin environments since it is destroyed by interstellar
radiation, e.g., CO has a lifetime of just $\sim 120$\,years.
Thus if a disk is observed to have a certain mass of CO, the rate at which CO is being replenished can be readily inferred.
Since observations of the dust disk can also be used to determine the rate at which the refractory component is being
replenished, these two rates can be used to find the volatile fraction of the planetesimals feeding the cascade.
While there are some assumptions in that calculation, such as that both gas and dust are being replenished from
steady state grinding, and the uncertainties in the CO mass measurement can be large if excitation conditions are not
known \citep[e.g.,][]{Matra2015}, this at least provides an order of magnitude estimate.

\begin{figure}
\hspace{1.4cm} \includegraphics[scale=0.50]{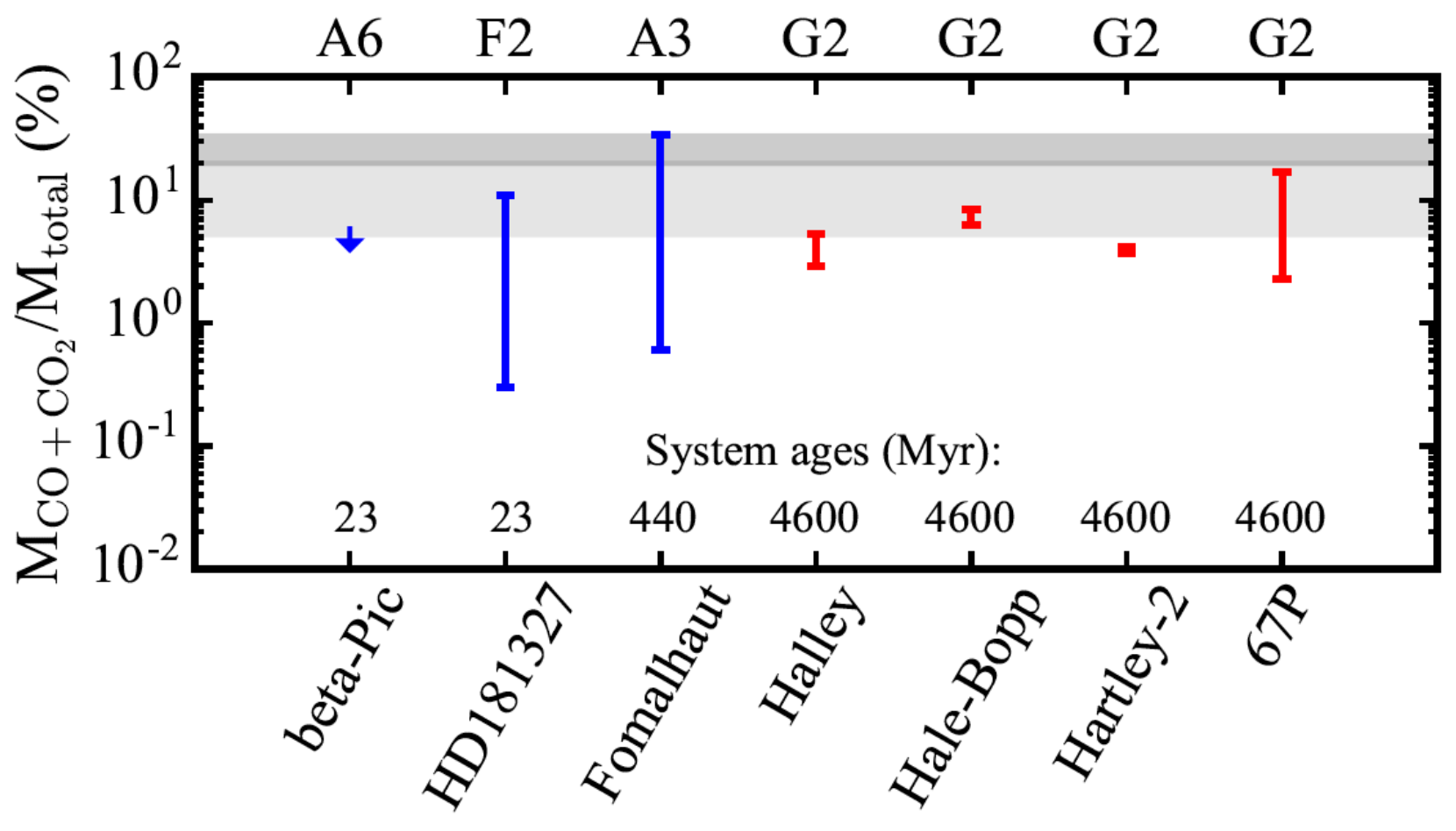}
\caption{Planetesimal volatile fraction inferred from observations of gas in debris disks compared with
the same fractions for Solar System comets \citep{Matra2017}.
Volatile fraction here is defined as the fraction of the total mass that is made up of CO and CO$_2$.}
\label{fig:4}
\end{figure}

For now this has been achieved for a handful of disks with conclusions being that the fraction of the planetesimals
that is made up of CO and CO$_2$ is similar to that of Solar System comets
\citep[see Fig.~\ref{fig:4},][]{Marino2016, Matra2017}.
Moreover the same model also explains those systems for which CO is not detected \citep{Kral2017}.
Thus it is plausible that all debris disks are made of volatile-rich planetesimals, and so that all have associated
secondary gas disks, but that only a fraction of these have CO at detectable levels.
While no other gas molecules have been detected in debris disks, this is to be expected given the lower abundance
relative to CO and moreover the even shorter lifetime of most molecules \citep{Matra2018a}. 

Atomic gas in the form of CI, CII and OI has also been detected \citep[e.g.,][]{RiviereMarichalar2014, Higuchi2017},
where it is inferred that this gas is the photodissociation product of CO and CO$_2$.
These are much longer lived and survive until the atomic gas disk has undergone viscous diffusion to be accreted
onto the star \citep{Kral2016}.
The observed ratio of carbon to CO thus constrains the viscous diffusion timescale, with implications for the
underlying physics of that accretion \citep{Kral2016b}.
For $\beta$ Pic, this resulted in a prediction for the amount of OI that was too low compared with that observed,
which was used to infer that the planetesimals were also water-rich, similar to Solar System comets, since that
would explain the detection \citep{Kral2016}.

While there is an emerging paradigm that has been outlined above, it is worth noting that there remain uncertainties.
For example, the model does not explain the distribution of CI observed in $\beta$ Pic \citep{Cataldi2018}.
It was also suggested that the $<50$\,Myr systems with protoplanetary disk levels of CO have a component of primordial
gas that has yet to disperse \citep{Kospal2013, Moor2017}.
However, it was recently shown that such high levels of CO could be explained in a secondary production scenario,
since if the gas production rate is particularly high, or the viscous diffusion timescale is too long, then
sufficient carbon can accumulate to shield the CO from photodissociating, which then can also start to self-shield
\citep{Kral2018}.

In consideration of the Solar System, the points to take away are that the volatile-rich composition of our Kuiper belt
may be similar for other stars, and also that there may have been a tenuous atomic gas disk present early on.

%%%%%%%%%%%%%%%%%%%%%%%%%%%%%%%%%%%%%%%%%%%%%%%%%%%%%
\subsection{Radial and Vertical Structure}
\label{ss:radial}

\begin{figure}
\includegraphics[scale=1.04]{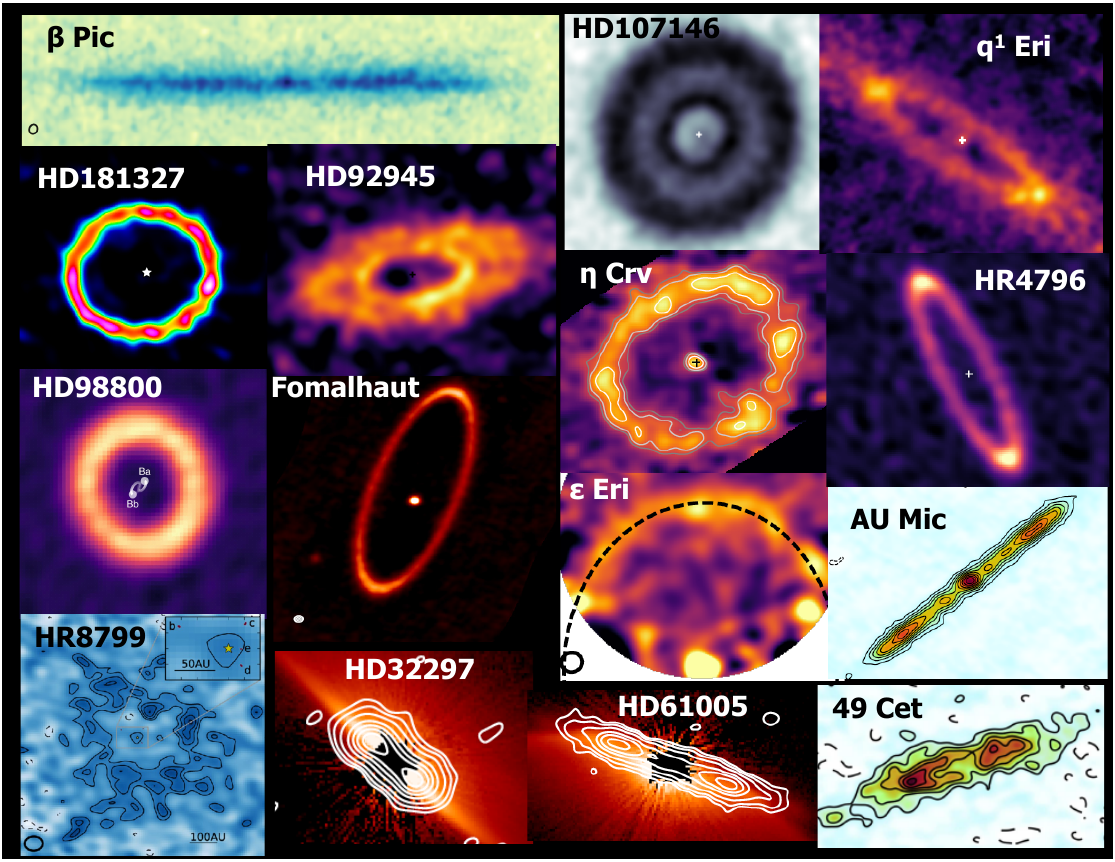}
\caption{Compilation of ALMA images of the distribution of mm-cm-sized dust in the debris disks of 15 nearby stars.
Individual images are from: $\beta$ Pic \citep{Matra2019b}, HD107146 \citep{Marino2018b}, q$^1$ Eri (J. Lovell, private communication),
HD181327 \citep{Marino2016}, HD92945 \citep{Marino2019}, $\eta$ Crv \citep{Marino2017a}, HR4796 \citep{Kennedy2018}, 
HD98800 \citep{Kennedy2019}, Fomalhaut \citep{MacGregor2017}, $\epsilon$ Eri \citep{Booth2017}, 
AU Mic \citep{Daley2019}, HR8799 \citep{BoothM2016}, HD32297 and HD61005 \citep{Schneider2014, MacGregor2018},
49 Cet \citep{Hughes2017}.
}
\label{fig:5}
\end{figure}

As Fig.~\ref{fig:1} shows, debris disks can be radially very narrow, with a fractional radial width (i.e., $\Delta r/r$)
of $\sim 10$\% in the example shown of the Fomalhaut disk.
That width can be measured for disks that have been imaged, but more specifically requires that disks
be resolved at mm wavelengths to
determine the width of the planetesimal belt, since the smaller dust probed at shorter wavelengths will be spread out
from the parent belt by radiation forces.
There is then some bias in the sense that a broad disk will have lower surface brightness than a narrow one, and so will
be harder to detect in high resolution imaging.
Nevertheless, Fig.~\ref{fig:5} shows that it is generally found that the disks are perhaps better described as rings or belts,
with the fractional radial width ranging from 10-100\%.

There are some notable exceptions to this description in that some disks are broad \citep[e.g.,][]{MacGregor2016}.
For example, the 61~Vir disk extends from 30\,au to beyond 150\,au
\citep{Wyatt2012, Marino2017b}, a property which means that its surface brightness
is so low that it is barely noticeable in ALMA imaging (but detected in the visibility data), with a radial profile that may be explained by
a broad disk of $\sim 10$\,km-sized planetesimals undergoing collisional erosion \citep{Marino2017b}.
Other examples of broad disks include HD107146 and HD92945, which also show an annular gap in the radial profile
\citep[see Fig.~\ref{fig:5},][]{Marino2018b, Marino2019} that may be related to planets either embedded in the disk or nearby
\citep{Yelverton2018}.

The inner edges of the debris belts have received some attention, since it is noted that if the belt is truncated by
dynamical interactions with a planet located at the inner edge, then the sharpness of the belt edge is indicative
of the mass of the planet \citep[e.g.,][]{Chiang2009}.
A more massive planet stirs up the disk to a greater extent resulting in a shallower slope.
Analogy with the Solar System shows that the sharpness and location of the disk's outer edge can also
be an important indicator of the system's past history \citep{Gladman2001}.

\begin{figure}
\includegraphics[scale=0.46]{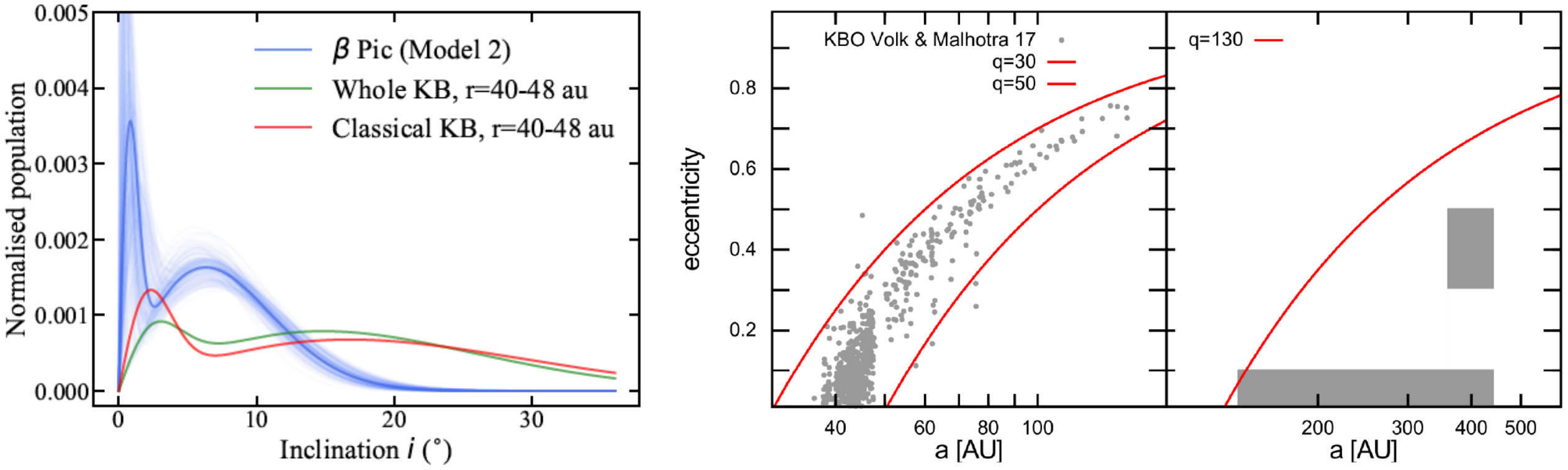}
\caption{Distributions of orbital elements inferred from ALMA images of debris disks that are analogous to dynamical
populations in the Kuiper belt.
{\it (Left)} Vertical structure in the $\beta$ Pic disk (see Fig.~\ref{fig:5}) is best modelled as the the sum of two
Gaussians that imply a distribution of inclinations reminiscent of the hot and cold classical Kuiper belt \citep{Matra2019b}.
{\it (Right)} Radial structure in the HR8799 disk (see Fig.~\ref{fig:5}) includes a halo of mm-sized grains that is
best modelled with a high eccentricity population potentially analogous to the scattered disk \citep{Geiler2019}.
}
\label{fig:6}
\end{figure}

The outer edge of a debris disk is sometimes quoted as the location at which the disk becomes too faint
to detect.
However, it is expected that for planetesimals confined to a belt, there would be
a halo of small $\mu$m-sized dust that extends beyond the belt due to radiation pressure \citep{Strubbe2006}.
In this case the outer edge of the planetesimal belt would be evident
as the location where there is a drop in the disk's surface brightness.
However, it is now becoming clear that
the extended dust components in some systems that were thought to be a halo of small dust
also contain mm-sized dust \citep{MacGregor2018}.
Since this dust is unlikely to be affected by radiation pressure, this calls into question whether
the planetesimal disk is in fact broad with simply a step change in density at what had been assumed to be the outer
edge of a radially confined planetesimal belt.
The presence of a faint outer belt of debris in one system could support this view \citep{Marino2016}. 
Alternatively the extended halo of mm-sized grains could represent an analogue of the scattered disk,
i.e., made of planetesimals (and the dust derived from their destruction) that are on eccentric orbits
undergoing scattering with a planet located near the inner edge.
Indeed the favoured interpretation of the radial structure of the HR8799 disk is that it has a low eccentricity disk
(i.e., a classical Kuiper belt analogue) as well as a high eccentricity population of comets in a scattered disk analogue
\citep[see Fig.~\ref{fig:6},][]{Geiler2019}, an interpretation strengthened by the fact that multiple planets interior to the belt are
known in this system \citep{Marois2010}.

It was already noted in \S \ref{ss:evolution} that the vertical structure of some disks has now been resolved.
Furthermore, for $\beta$ Pic it was found that the vertical structure was not well described by a single Gaussian, rather it
is fit well as the sum of two Gaussians that correspond to components with mean inclinations of $1^\circ$ and $9^\circ$
\citep{Matra2019b}.
While this is a few times flatter than the Solar System, there is a clear analogy with the cold and hot components
of the classical Kuiper belt (see Fig.~\ref{fig:6}).

%%%%%%%%%%%%%%%%%%%%%%%%%%%%%%%%%%%%%%%%%%%%%%%%%%%%%
\subsection{Dynamical Structures}
\label{ss:structures}
Some of the radial structures discussed in \S \ref{ss:radial} can be considered dynamical structures, in that
the shape of inner and outer disk edges, and gaps within the disks, can all be related to the presence of
planets.
However, such an interpretation is potentially ambiguous, because this may alternatively be indicative of the initial
distribution of planetesimals, in which radial structure may have been imprinted by processes ongoing
during the protoplanetary disk phase (e.g., the formation of planetesimals at ice-lines);
i.e., no planets may be required.
Non-axisymmetric structure is a clearer indicator that additional perturbations are ongoing, since these
would be needed to maintain such structures over long timescales.

Several of the disks that have been imaged are seen to be eccentric.
This was manifested originally by a brightness asymmetry in mid-IR images of the HR4796 disk \citep{Telesco2000},
which was interpreted as arising from the pericentre side of the disk being closer to the star and so
hotter and brighter \citep{Wyatt1999}.
That interpretation is supported by ALMA imaging which showed that the opposite side of the disk is brighter
\citep{Kennedy2018} as expected due to the slower orbital velocity at apocentre \citep{Pan2016}.
Subsequently scattered light imaging showed that the star is not exactly at the centre of the disk
\citep{Milli2017}, although the direction of the offset is not exactly that expected suggesting that
there may be some additional density variations around the ring \citep[see also][]{Lohne2017}. 
Fomalhaut's disk (see Fig.~\ref{fig:1}) is perhaps a better example, being brighter and larger on the sky so that
the offset and asymmetries are more clearly seen \citep{Kalas2005, MacGregor2017}, with additional azimuthal structure
also evident \citep{Kalas2013}.

The origin of these eccentricities, which are of order 0.05-0.15, have not been identified, but would arise naturally
from the secular perturbations of a planet on an eccentric orbit. 
While a planet candidate was detected orbiting near the inner edge of
the Fomalhaut disk \citep[see Fig.~\ref{fig:1},][]{Kalas2008} its eccentricity
was subsequently found to put it on a disk-crossing orbit \citep{Kalas2013} that would significantly disrupt the disk
suggesting that the planet is less massive than originally thought and/or has only been put on this orbit
recently \citep{Beust2014, Tamayo2014, Lawler2015}. 
The presence of an additional planet in the system that is responsible for the eccentricity is thus inferred.

A more secure identification of a disk asymmetry that is caused by planetary perturbations is the warp in the
$\beta$ Pic disk \citep{Mouillet1997}.
The change in the plane of symmetry seen in scattered light images at 80\,au in this edge-on disk \citep[see][]{Apai2015}
was inferred to arise from the secular perturbations of a planet on an inclined orbit at $\sim 10$\,au.
Since that planet has subsequently been found with direct imaging \citep{Lagrange2010}, this supports the idea that
disk structures can be used to identify unseen planets.
That said, there remains some uncertainty in the interpretation of the vertical structure of the $\beta$ Pic disk,
not least because the additional clumpy structure mentioned below is situated at the same location as the warp.

\begin{figure}
\hspace{2cm} \includegraphics[scale=0.60]{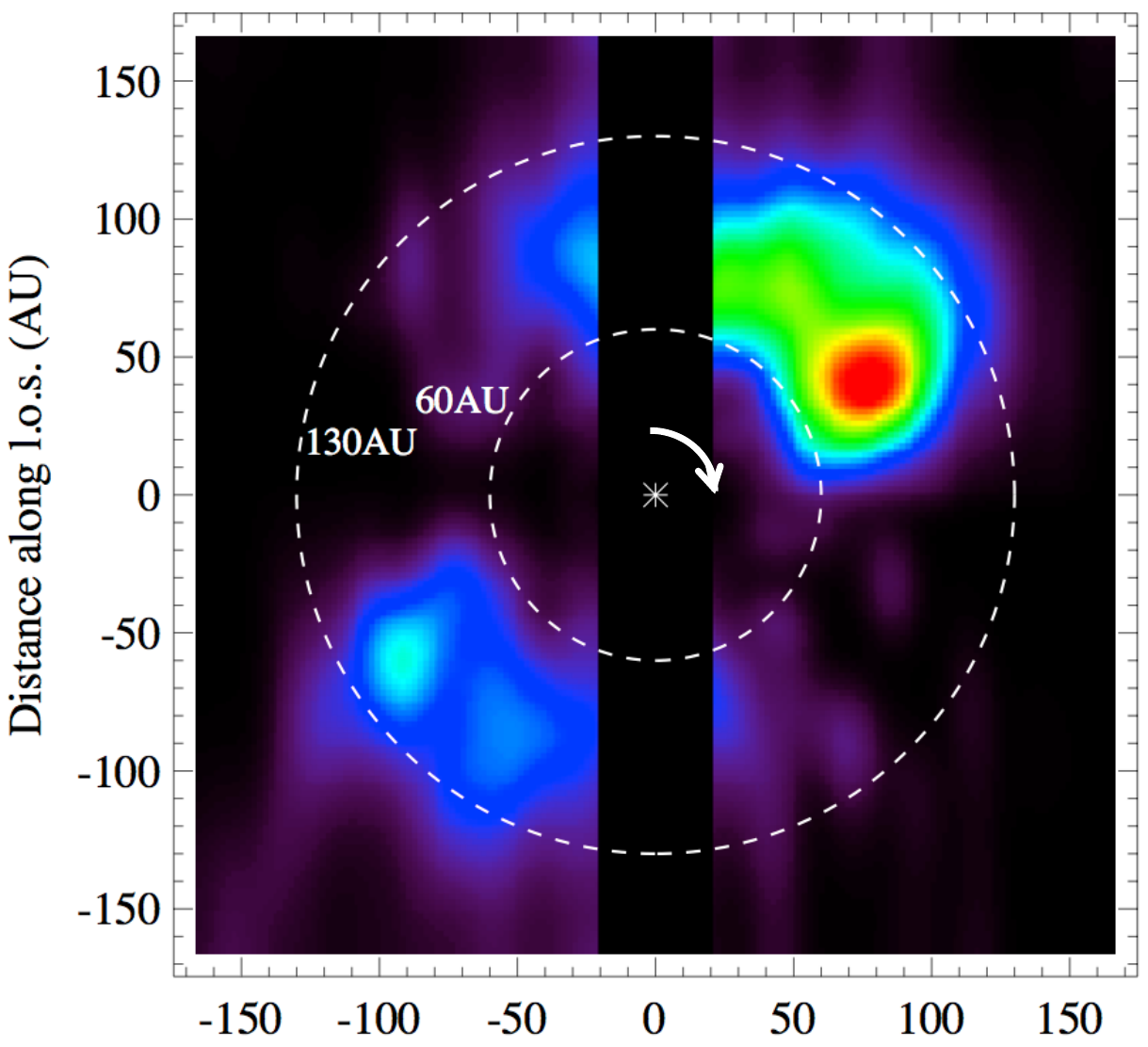}
\caption{Dynamical structure in the debris disk of $\beta$ Pic that is analogous to that of resonant Kuiper belt
objects.
While the disk is edge-on, velocity information from CO observations allows its face-on distribution to be
inferred showing that this includes two clumps of unequal brightness on opposite sides of the star \citep{Dent2014}.
The best explanation is that outward migration of a planet close to the inner edge of the disk trapped volatile-rich
planetesimals into its 2:1 and 3:2 resonances causing over-densities in these regions where collisions are
most frequent \citep{Wyatt2003}.
The CO lifetime is short compared with the orbital timescale meaning that it does not travel far from where
it is created before being photodissociated.
}
\label{fig:7}
\end{figure}

Both the Vega and $\epsilon$~Eri debris disks were inferred to be clumpy based on sub-mm imaging
\citep{Holland1998, Greaves1998}.
Subsequent imaging gave conflicting views on the presence of the clumps \citep{Greaves2005, Marsh2006, Hughes2012, Holland2017},
but it seems likely that at least some of the clumpiness was attributable to background galaxies.
The most secure evidence for clumpiness in a disk comes from $\beta$ Pic for which mid-IR images show a prominent
clump at $\sim 50$\,au in projected separation from the star \citep{Telesco2005}.
The same clump was later imaged in CO \citep[see Fig.~\ref{fig:7},][]{Dent2014}, for which the additional kinematic information showed that
the clump is both radially and azimuthally broad and situated at $\sim 85$\,au distance from the star \citep{Matra2017b}.
Two models were proposed to explain the clump, either a recent collision between two Mars-sized protoplanets
\citep{Jackson2014}, or the trapping of planetesimals into the 2:1 and 3:2 resonances of a $\sim 30M_\oplus$ planet
that migrated out from 31-57\,au over 23\,Myr \citep{Wyatt2003, Matra2019b}.
The radial breadth of the clump disfavours a collisional origin leaving the favoured interpretation as a structure
analogous to the resonant Kuiper belt objects (the Plutinos and Twotinos).
Comparison of Figs.~\ref{fig:2} and \ref{fig:7} show there are some notable differences,
however, such as that the tight clump observed in the $\beta$ Pic disk would require the planetesimals to have
a smaller libration width than resonant objects in the Kuiper belt.
This could be explained within the context of this model
by the stochasticity of the migration and/or the initial orbital eccentricities \citep[e.g.,][]{Reche2008, Nesvorny2016}.

%%%%%%%%%%%%%%%%%%%%%%%%%%%%%%%%%%%%%%%%%%%%%%%%%%%%%
\subsection{Connection from Outer to Inner System}
\label{ss:connection}
The fraction of stars with excess 10\,$\mu$m emission indicative of warm dust in the region where there
may be habitable planets was long recognised to be low \citep{Aumann1991}, and was quantified more recently
from the WISE photometric survey to find that excesses above 10\% of the stellar emission only
occur to 1:$10^4$ nearby stars \citep{Kennedy2013},
although this rate is higher if considering a sample of young $<100$\,Myr stars. 
This result is explained by collisional evolution \citep[see also Fig.~\ref{fig:3} and \S \ref{ss:mass},][]{Wyatt2007a},
which suggests that the rare bright excesses are transient populations of warm dust, perhaps originating in recent
collisions.
However, ongoing interferometric surveys have shown that excesses at 0.1-1\% levels are more common, and present
to $\sim 20$\% of stars \citep{Ertel2018}, and moreover are more prevalent (though not exclusively) around
stars known to host cold outer disks \citep{Mennesson2014}.

This does not prove that the warm dust must originate somehow in the outer belt, since it could be that the
conditions of a protoplanetary disk that leads to the formation of an outer cold belt also favour the formation
of a massive asteroid belt \citep[e.g.,][]{Geiler2017}.
However, there are two mechanisms by which outer belts can replenish a warm dust population.
The first is by Poynting-Robertson drag \citep{Burns1979}.
While the dust population dragged in from the known outer belts is expected to be significantly depleted by mutual
collisions \citep[which is not the case in the more tenuous belt in the Solar System,][]{Wyatt2005},
the levels of dust expected to make it in to $\sim 1$\,au are comparable to those inferred from
observations in some cases \citep{Mennesson2014, Kennedy2015},
although the dust population may be further depleted if the dust has to cross the path of any planets
\citep{Bonsor2018}.
The second is by comet scattering.
That is, an intervening planetary system could scatter planetesimals into the inner regions where they could
sublimate or disintegrate to replenish the warm dust, much as comets in the Solar System replenish the zodiacal
cloud \citep{Nesvorny2010}.

\begin{figure}
\includegraphics[scale=0.66]{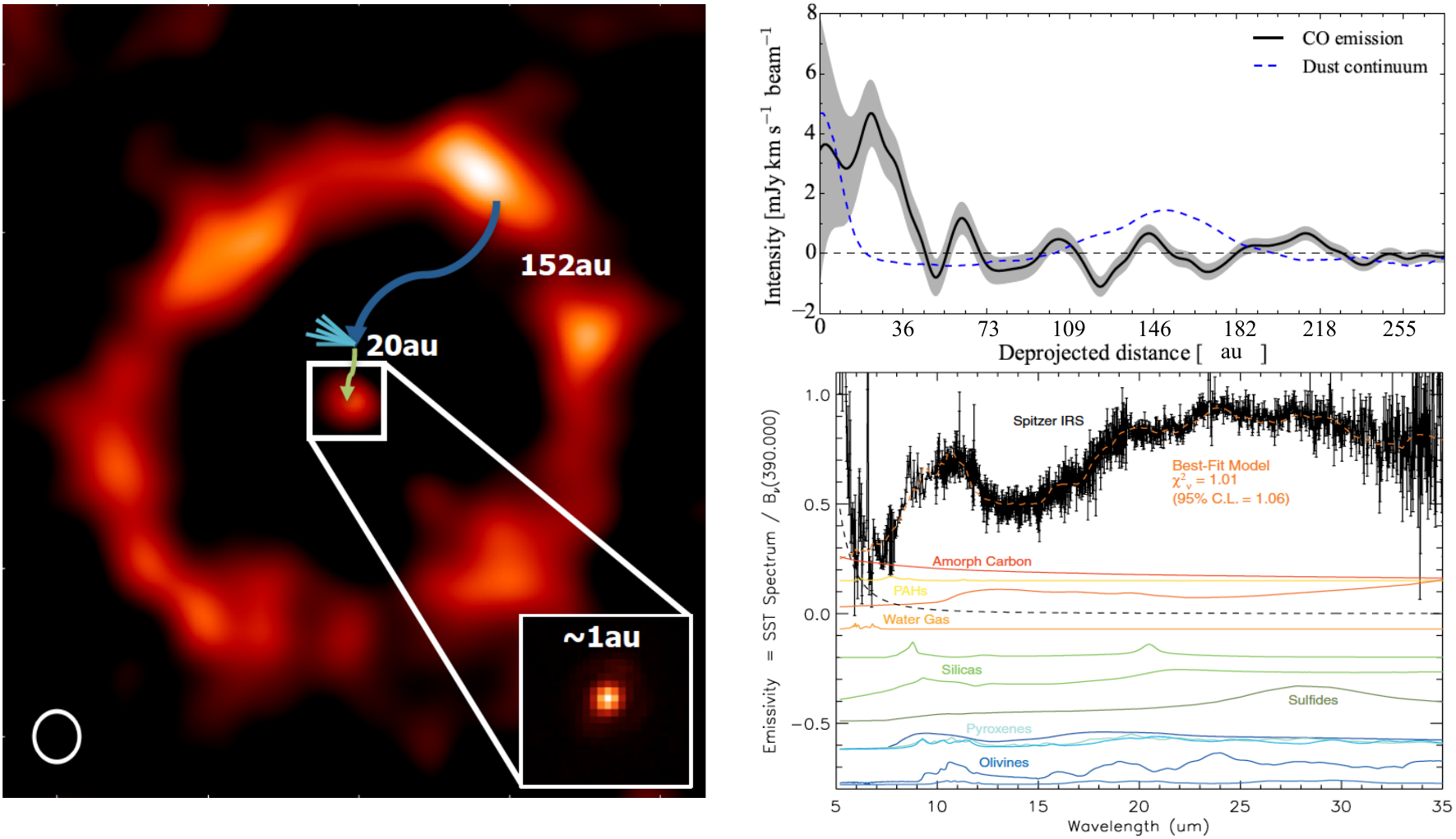}
\caption{Evidence for inward scattering of exocomets in the debris disk of the $\sim 1$\,Gyr F star $\eta$ Crv.
{\it (Left)} ALMA image of its outer debris belt
(and the stellar photospheric emission at the centre),
with inset (bottom right) of the mid-IR emission showing warm dust close to the
star, and annotation to illustrate the origin of that warm dust in exocomet scattering \citep{Marino2017a}.
{\it (Top right)} The profile of CO emission showing this peaks at $\sim 20$\,au close to where the exocomets 
may be expected to cross the H$_2$O or CO$_2$ ice lines \citep{Marino2017a}.
{\it (Bottom right)} The spectrum of mid-IR emission can be fitted by a model for the dust composition that
suggests it is water- and carbon-rich \citep{Lisse2012}.
}
\label{fig:8}
\end{figure}

For $\eta$ Crv shown in Fig.~\ref{fig:8}, its $\sim 10$\% excess from dust at
$\sim 1$\,au \citep{Smith2009, Defrere2015}
cannot be explained by steady state collisions in an asteroid belt given its $\sim 1$\,Gyr age \citep{Wyatt2007a}.
In this case an origin in the outer belt seen at 150\,au \citep{Wyatt2005b} is most likely, and is supported by the
composition of the warm dust inferred from its infrared spectrum \citep{Lisse2012}.
The warm dust is too bright to be explained by P-R drag, but inward scattering of comets is the favoured
scenario, particularly following the tentative detection of CO gas at $\sim 20$\,au from the star
\citep{Marino2017a}.
The detection of gas at a location where there is no dust is surprising, but is readily explained if this is the
location where comets start sublimating on their way in before depositing their dust at $\sim 1$\,au.

The requirement to sustain a high rate of comet scattering sets constraints on the intervening planetary system, 
e.g. since the comets should not be ejected before reaching the inner regions \citep{Wyatt2017}, but it is found
that a plausible planetary system architecture is one with a chain of $3-30M_\oplus$ planets \citep{Marino2017a}.
More generally, some level of comet scattering is expected in all systems with an outer belt, with the
rate determined by the processes that perturb planetesimals out of their orbits in the belt, and the
architecture of their planetary system \citep{Marino2018a}.
Since these planetesimals may impact any planets and so deliver water or strip their atmospheres,
this has significant interest for those seeking to understand the prevalence of conditions that favour the
development of life.

%%%%%%%%%%%%%%%%%%%%%%%%%%%%%%%%%%%%%%%%%%%%%%%%%%%%%
%%%%%%%%%%%%%%%%%%%%%%%%%%%%%%%%%%%%%%%%%%%%%%%%%%%%%
\section{Conclusions}
\label{s:conclusions}
The Sun is not unique in having either a planetary system or in having belts of planetesimals. 
Other stars also have what can be considered as extrasolar analogues to the Kuiper belt.
However, the extent of that analogy depends on how strict the similarities must be.
Certainly other stars have belts of planetesimals at distances from their stars similar to our own
Kuiper belt, but these belts contain orders of magnitude more mass, perhaps comparable to
that in our primordial Kuiper belt.
Since we cannot yet detect a true Kuiper belt analogue around another star
we cannot yet tell how ours compares within the population, except to say that it is not in
the top 20\%.

There are, however, many other similarities to the Kuiper belt in the debris disks observed
around other stars.
For example, several have volatile-rich compositions inferred to be similar to Solar System
comets (from gas observations).
In terms of structures, there is evidence for analogues to the hot and cold classical belt
(from vertically resolved edge-on disks), to the resonant Kuiper belt object population (from clumpy disks),
and to the scattered disk (from haloes of mm-sized grains).
Many disks are seen to be radially narrow with sharp inner and outer edges (from resolved imaging).
It is also inferred that in some systems exocomets are scattered into the inner regions
(from observations of close-in dust and gas).

It is nevertheless worth bearing in mind that these similarities are based in some cases on
a relatively small number of what must inevitably be considered the extremes of the debris disk
population (in that they are bright enough to have been studied in sufficient detail to identify
any similarity).
Also, the similarities that are highlighted in the literature (and in this review)
reflect to some extent our inherently anthropocentric view (e.g., some of the aforementioned
features may have other interpretations).
Moreover, there are also examples of disks that do not fit into a Kuiper belt-like
mold (e.g., the broad disks, or the eccentric rings). 

These differences in the extrasolar debris disk population are to be expected, because
just as we have learned by comparing the Solar System to
the exoplanet population, planet formation has a diverse range of outcomes of which
the Solar System is just one example.
By studying and comparing our Kuiper belt to extrasolar debris disks, we have
a different perspective on those outcomes, one which is biased towards processes that
occur in the outer regions of the systems that are hard to access through
exoplanet observations.
What the similarities do show is that, perhaps, we share some dynamical processes
in common with other stars, and by considering the Solar System as one outcome we
may achieve a deeper understanding of the planet formation process and the uniqueness of our own situation.

%%%%%%%%%%%%%%%%%%%%%%%%%%%%%%%%%%%%%%%%%%%%%%
%%%%%%%%%%%%%%%%%%%%%%%%%%%%%%%%%%%%%%%%%%%%%%
%%%%%%%%%%%%%%%%%%%%%%%%%%%%%%%%%%%%%%%%%%%%%%
\bibliographystyle{spbasicHBexo}
\bibliography{wyattbib}

\end{document}